\documentclass{emulateapj}
\newcommand{\mach}{\mathcal{M}}
\newcommand{\anonlin}{\mathcal{A}}
\newcommand{\zz}{\mbox{\boldmath $z$} {}}
\newcommand{\xx}{\mbox{\boldmath $x$} {}}
\begin{document}
\title{The gravitational drag force on an extended object moving in a gas}
\author{Cristian G.~Bernal and F.~J.~S\'anchez-Salcedo}
\affil{Instituto de Astronom\'{\i}a, Universidad Nacional Aut\'onoma
de M\'exico,
P.O. Box 70-264, Ciudad Universitaria, 04510, Mexico City, Mexico }

\begin{abstract}
Using axisymmetrical numerical simulations, we revisit the gravitational drag
felt by a gravitational Plummer sphere with mass $M$ and core radius $R_{s}$, moving
at constant velocity $V_{0}$
through a background homogeneous medium of adiabatic gas. Since the potential is non-diverging,
there is no gas removal due to accretion.
When $R_{s}$ is larger than the Bondi radius $R_{B}$, the perturbation is linear at
every point and the drag force is well fitted by the time-dependent Ostriker's formula
with $r_{\rm min}= 2.25R_{s}$, where $r_{\rm min}$ is the minimum impact parameter in
the Coulomb logarithm.  In the deep nonlinear supersonic regime ($R_{s}\ll R_{B}$), the minimum
radius is no longer related with $R_{s}$ but with $R_{B}$. We find  $r_{\rm min}=3.3\mach^{-2.5}R_{B}$,
for Mach numbers of the perturber between $1.5$ and $4$, although $r_{\rm min} = 2\mach^{-2}R_{B}=2GM/V^{2}_{0}$ also
provides a good fit at $\mach>2$.  As a consequence, the drag force does not depend sensitively
on the nonlinearity parameter $\anonlin$, defined as $R_{B}/R_{s}$, for $\anonlin$-values larger
than a certain critical value $\anonlin_{\rm cr}$. 
We show that our generalized Ostriker's formula  for the drag force is more accurate than the 
formula suggested by Kim \& Kim (2009).

\end{abstract}

\keywords{interstellar medium -- ISM: structure --
 hydrodynamics}


\section{Introduction}
\label{sec:introduction}
A gravitational perturber moving through a gaseous background
creates a density wake in the medium.  The complexity of the wake
has motivated a number of groups to tackle the problem using
numerical simulations (e.g., Hunt 1971, 1979; Shima et al.~1985; Ruffert 1994, 1996;
S\'anchez-Salcedo \& Brandenburg 2001; Cant\'o et al. 2011, 2012).
The gravitating object induces  small disturbances in the far-field
ambient medium (i.e. in streamlines with large impact parameters) and,
consequently, the far-field density structure of the wake can be derived 
in linear perturbation theory  (Dokuchaev 1964; Ruderman \& Spiegel 1971).
Using this approach, Ostriker (1999) derived the density wake behind a gravitational
body of mass $M$ moving at velocity $V_{0}$ on a straight-line
trajectory through a homogeneous infinite
medium with unperturbed density  $\rho_{\infty}$ and sound speed
$c_{\infty}$, when the perturber is dropped at $t=0$.
For subsonic perturbers, the isodensity contours are closed ellipsoids,
which do not contribute to the drag force, except in the outer parts of
the wake where the ellipsoids are not closed. Supersonic perturbers
generate a density wake only within the rear Mach cone. 

Here, we are interested in the dynamical friction, that is, the drag
force onto the perturber due to the gravitational interaction with
its own induced wake.
Thus, once the density of the wake is known, the drag force $F_{DF}$ can be computed
as the gravitational force between the perturber and its wake.
The drag force is usually written as
\begin{equation}
F_{DF}=\frac{4\pi \rho_{\infty} (GM)^{2}}{V_{0}^{2}}\ln\Lambda,
\label{eq:rephaeli}
\end{equation}
where $\ln\Lambda$ is the Coulomb logarithm. For subsonic
perturbers, i.e. $\mach\equiv V_{0}/c_{\infty} <1$, Ostriker (1999)
found that, at times satisfying $c_{\infty}t>r_{\rm min}/(1-\mach)$,
the Coulomb logarithm is
\begin{equation}
\ln\Lambda=\frac{1}{2}\ln \left(\frac{1+\mach}{1-\mach}\right)
-\mach,
\label{eq:ostriker1}
\end{equation}
where $r_{\rm min}$ is
the minimum radius of the effective gravitational interaction of
a perturber with the gas. In practice, for a point mass object, 
$r_{\rm min}$ is taken as the radius
at which the linear approximation breaks down. 
For supersonic perturbers, the Coulomb logarithm is given by
\begin{equation}
\ln\Lambda=\frac{1}{2}\ln \left(1-\mach^{-2}\right)+
\ln\left(\frac{\mach c_{\infty}t}{r_{\rm min}}\right),
\label{eq:ostriker2}
\end{equation}
for $\mach>1$ and $c_{\infty}t>r_{\rm min}/(\mach-1)$ (Ostriker 1999).

S\'anchez-Salcedo \& Brandenburg (1999) tested numerically Ostriker's
formula when the perturber is extended. This could represent a globular cluster 
orbiting in the gaseous halo of its host galaxy, a young massive star cluster
in a merging system, an elliptical galaxy depleted 
of gas in the intracluster medium, or a galaxy cluster in a major
merger environment (e.g., Naiman et al. 2011). S\'anchez-Salcedo \& Brandenburg (1999) 
found that Ostriker's formula
describes successfully the temporal evolution and magnitude of the force experienced by
a Plummer perturber,  when its mass $M$
is small enough so that the gas response to the perturbation is linear. 
For supersonic perturbers, they inferred $r_{\rm min}=2.25R_{s}$,
where $R_{s}$ is the core radius.

Kim \& Kim (2009, hereafter KK09) carried out axisymmetric numerical simulations
to study the flow past a Plummer sphere in the linear 
and in the nonlinear regimes.
For the linear supersonic regime,
they found that the minimum radius in Eq. (\ref{eq:ostriker2}) depends slightly on the Mach
number: $r_{\rm min}=0.35{\mathcal{M}}^{0.6}R_{s}$. Moreover, they found in their
simulations that the drag force is highly suppressed in the nonlinear regime as compared
to the linear case.

In this paper, we revisit the drag force on an extended (non-accreting) gravitating 
object with a range of velocities relative to the ambient medium. Our aim is to provide 
a physically more motivated and more
accurate formula for the drag force in both the linear and nonlinear
cases. 
In Section \ref{sec:radius}, we briefly recapitulate the relevant scales in
the problem, discuss the ambiguity in the definition of the minimum
radius and describe the numerical code adopted in this paper.
Section \ref{sec:linearcase} is devoted to discuss the drag force in the linear
case. The results for nonlinear simulations are analyzed in Section \ref{sec:nonlinear}.
Finally, we summarize our findings in Section \ref{sec:conclusions}.

\section{ Dynamical friction in a gas: Relevant scales, 
the minimum cut-off distance and the inner boundary condition}
\label{sec:radius}
Consider the flow pattern past a gravitational object.
The inner structure of the wake (in the vicinity of the object) depends on the adopted
equation of state for the gas (e.g., $\gamma$, if the gas behaves as a polytrope, 
$P\propto \rho^{\gamma}$) and depends also
on  whether the perturber
is a point mass gravitational potential or a non-point mass distribution (e.g., Naiman
et al. 2011). As a consequence, the drag force may depend on $\gamma$
and on the adopted inner conditions. For instance,
Lee \& Stahler (2011) found that,  for subsonic perturbers with $\mach<0.75$,
the drag force felt by a point mass (accretor) is higher than what
Ostriker's formula predicts for extended non-accretors. 
For hypersonic perturbers, however,
Cant\'o et al.~(2011) argued that the drag force expressions (\ref{eq:rephaeli}) and (\ref{eq:ostriker2}), 
are still valid for a point mass accretor, provided that $r_{\rm min}$
is chosen appropriately. Cant\'o et al.~(2011) demonstrated that
$r_{\rm min}\simeq 0.8GM/V_{0}^{2}$ for a hypersonic point mass.

\begin{figure}
  \includegraphics[width=85mm,height=104mm]{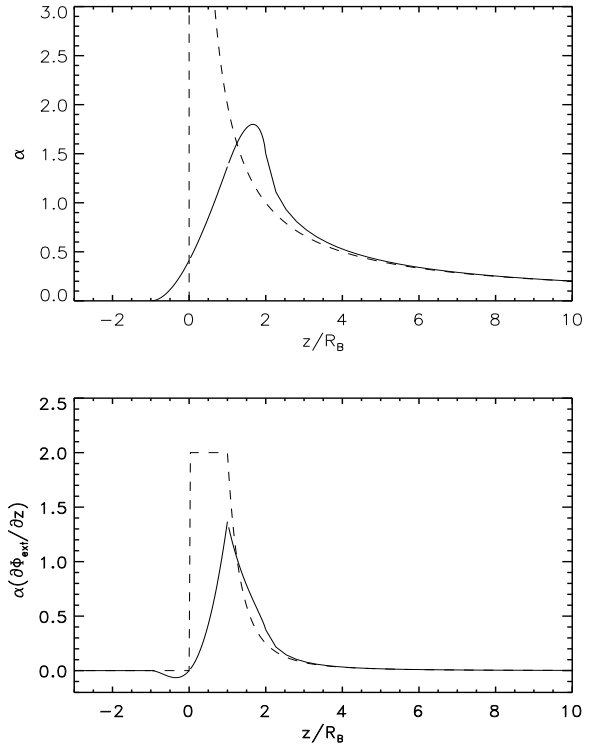}
  \caption{Cuts along the symmetry axis of the perturbed density 
$\alpha\equiv (\rho-\rho_{\infty})/\rho_{\infty}$ 
(upper panel) and of the magnitude $\alpha \partial\Phi_{\rm ext}/\partial z$ 
in units of $GM/R_{B}^{2}$ (lower panel), using linear theory,
for a perturber of mass $M$ traveling at Mach $2$. 
To compute $\alpha$, we consider two cases for the pertuber: a point-mass 
perturber (dashed line) and a spherical perturber of constant density 
with a radius $a=R_{B}$ (solid line). 
In order to illustrate the error made when Equation (\ref{eq:fdfapprox}) 
is used instead of Equation (\ref{eq:fdfsimple}),
we plot the term $\alpha \partial\Phi_{\rm ext}/\partial z$, 
which appears in the integrand of these equations,
where $\Phi_{\rm ext}$ represents the gravitational potential created 
by a homogeneous sphere of radius $R_{B}$. 
}
  \label{fig:alpha_sphere}
\end{figure}

For a non-point mass perturber,
there are three independent characteristic radii in the problem: the softening length $R_{s}$,
the Bondi radius $R_{B}\equiv GM/c_{\infty}^{2}$ and the gravitational radius or the so-called
accretion radius\footnote{Note that our potential is non-diverging and mass is conserved, 
implying that there is no accretion but accumulation of mass around the perturber.
Still, we use the term ``accretion radius'' to refer to $R_{A}$.}
$R_{A}\equiv 2GM/(c_{\infty}^{2}+V_{0}^{2})=2R_{B}/(1+{\mathcal{M}}^{2})$.
On dimensional grounds, one expects that
the minimum radius $r_{\rm min}$ in Equation (\ref{eq:ostriker2}) may be expressed
as a certain combination of the three characteristic radii.

The linear analysis is expected to provide a good estimate of the drag force
in circumstances where the depth of the external gravitational potential 
$\Phi_{\rm ext}(0)$ is small as compared to $c_{\infty}^{2}$, so that the disturbances
are small at any location. 
Therefore, the linear analysis is tacitly assuming that
the gravitational potential is non-divergent. 
For an extended perturber with softening radius $R_{s}$,
the perturbation is linear at any location if the parameter 
${\mathcal{A}}\equiv\frac{GM}{c_{\infty}^{2}R_{s}}=R_{B}/R_{s}\ll 1$. 
In this situation, 
the only relevant characteristic radius is $R_{s}$ and, hence,
one expects a relationship between $r_{\rm min}$ and $R_{s}$: 
$r_{\rm min}=\lambda({\mathcal{M}}) R_{s}$. In this case, the response of
the fluid is linear at any position and, therefore, the drag force is expected to vary with
$\mach$ according to  Equations (\ref{eq:rephaeli})-(\ref{eq:ostriker2}).
In other words, $\lambda$ is expected to depend very
weakly on the Mach number in order to preserve the functional dependence
with $\mach$.  The drag force depends implicitly on $\gamma$ through the 
sound speed in the medium. This suggests that  $\lambda\simeq \lambda_{0}$, being
$\lambda_{0}$ a Mach-independent constant. 

Consider now the nonlinear case where $R_{B}\gg R_{s}$ 
(i.e. ${\mathcal{A}}\gg 1$). Under this circumstance,  
the softening radius is likely irrelevant
and thus one expects $r_{\rm min}$ to be linked to either $R_{B}$ or
$R_{A}$ (or both) in this limit. Since $R_{A}$ and $R_{B}$ are simply related through
$R_{A}=2R_{B}/(1+\mach^{2})$, it is rather general to assume that
$r_{\rm min}=\tilde{\lambda}({\mathcal{M}};\gamma)R_{B}$. 
In order to estimate $\tilde{\lambda}$, we need to study the
behaviour of the flow pattern within the Bondi radius.

\begin{figure}
  \includegraphics[width=85mm,height=80mm]{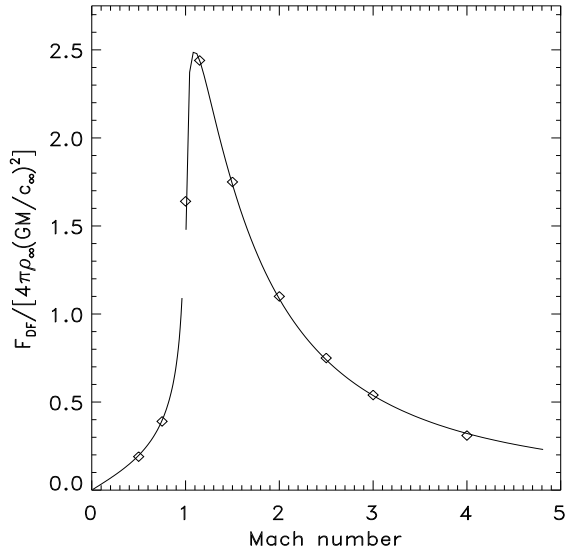}
  \caption{Dimensionless drag force at $t=100R_{s}/c_{\infty}$
as a function of Mach number (defined
as $V_{0}/c_{\infty}$) for a Plummer perturber in the linear regime ($\anonlin=0.01$).
The solid line corresponds to the Ostriker's formula (Eqs. \ref{eq:ostriker1}-\ref{eq:ostriker2})
with $r_{\rm min}=2.25R_{s}$.
  }
  \label{fig:force_mach_lin}
\end{figure}

Our aim is to find $\lambda({\mathcal{M}})$ and $\tilde{\lambda}({\mathcal{M}};\gamma)$
for a gravitational perturber, which is described by a Plummer model, in an adiabatic
gas ($\gamma=5/3$).
To that end, we have carried out high-resolution axisymmetric simulations with
a customized version of the hydrodynamic code FLASH4.0. 
This scheme and tests of the code are described in Fryxell et al. (2000).
In particular, we use the FLASH UG method (Uniform Grid) with the split $8$-wave
solver to solve the whole set of HD equations in cylindrical coordinates. Our grid
models have $10000\times 5000$ zones in $(R,z)$, with
$6$ zones per $R_{s}$, to match the convergent requirement given in
KK09. We further tested convergence of our models for
several resolutions and domain sizes. Our box is large enough for the wake not to
reach any of the boundaries of the domain.

\section{The linear case: The values of $\lambda$  }
\label{sec:linearcase}
We consider a gravitational perturbing body of mass $M$ at the origin of our coordinate
system, surrounded by a gas whose velocity far from the particle
is $\mach c_{s}\hat{\zz}$. Suppose that the perturber is turned on at $t=0$.
Denote by $\rho_{\rm ext}$ the specified
density field of the perturber and by $\Phi_{\rm ext}$ its gravitational
potential, thus $\nabla^{2}\Phi_{\rm ext}=4\pi G \rho_{\rm ext}$.
As already discussed in \S \ref{sec:radius}, the introduction of a softening radius
in the gravitational potential allows us to have cases where the response
of the gas to the perturbation is linear at any position in space.
This occurs when ${\mathcal{A}}\ll 1$. Then, the perturbed density,
defined as  $\alpha=(\rho-\rho_{\infty})/\rho_{\infty}$,
can be calculated at any point in space by 
\begin{equation}
\alpha(\xx,t)=\frac{1}{M} \int \tilde{\alpha}(\xx-\xx',t)\rho_{\rm ext}(\xx') d^{3}x',
\label{eq:alpha_extended}
\end{equation}
where 
\begin{equation}
\tilde{\alpha}(R,z,t)=\frac{\xi GM}{c_{\infty}^{2}}
\frac{1}{\sqrt{z^{2}-R^{2}(\mach^{2}-1)}}.
\label{eq:alpha_axisymmetric}
\end{equation}
Here, $z$ is in the direction of motion, 
$R=\sqrt{x^2+y^2}$ is the cylindrical radius and
 \[ \xi = \left\{ \begin{array}{ll}
         1 & \mbox{if $R^{2}+z^{2}<(c_{\infty}t)^{2}$};\\
         2 & \mbox{if ${\mathcal{M}}>1$, $R^{2}+z^{2}>(c_{\infty}t)^{2}$, $z/R>(\mach^{2}-1)^{1/2}$} \\
             & \mbox{{\rm and}\,\,$z<(\mach^{2}-1)c_{\infty}t/\mach$} ;\\
         0 & \mbox{otherwise,} \end{array} \right. \]
(see, e.g., Just \& Kegel 1990; Furlanetto \& Loeb 2002; KK09).

In order to illustrate how the density in the wake depends
on  the size of the perturber, consider
the simplest case where the perturber is a sphere of constant density
with radius $a$, that is 
$\rho_{\rm ext}=3M/(4\pi a^{3})$ at $r<a$, and $\rho_{\rm ext}=0$ at $r>a$.
Note that, in this model, the gravitational force that the gas feels 
at $r>a$ is identical to the force created by a point mass.
In Figure \ref{fig:alpha_sphere}, we plot the perturbed density
$\alpha$ along the $z$-axis for $a=R_{B}$ and $\mach=2$, in
the stationary wake.
For comparison, we also plot the perturbed density derived in linear theory
for a point mass. We see that the perturbed
density $\alpha$ is no longer singular at the origin ($r\rightarrow 0$) 
when $a\neq 0$.  Indeed the maximum
of $\alpha$ occurs at $z=1.75a$ (note that $a=R_{B}$ in Figure
\ref{fig:alpha_sphere}).
We also see that $\alpha\simeq \tilde{\alpha}$ at $z\simeq 1.25a$ and
at large enough distances $z\gtrsim 3a$.

Once $\alpha(R,z,t)$ is known, the strength of drag force can be derived as:
\begin{equation}
F_{DF}=2\pi\rho_{\infty}\int \alpha \frac{\partial\Phi_{\rm ext}}{\partial z} R\, dR \,dz.
\label{eq:fdfsimple}
\end{equation}
In a general case, the volume integral (\ref{eq:fdfsimple}) must be computed
numerically.  As a rough first order approximation, however,
Rein (2011) estimated the minimum cut-off radius $r_{\rm min}$ by
approximating the drag force by
\begin{equation}
F_{\rm approx}=2\pi\rho_{\infty}\int \tilde{\alpha} \frac{\partial\Phi_{\rm ext}}{\partial z} R\, dR \,dz,
\label{eq:fdfapprox}
\end{equation}
that is, by replacing $\alpha$ for $\tilde{\alpha}$ in Equation (\ref{eq:fdfsimple}).
The reason is that at those distances 
where $\alpha$ is dissimilar to $\tilde{\alpha}$,  it
holds that $\partial\Phi_{\rm ext}/\partial z$ is small. For instance, when the
perturber is described by a homogeneous sphere, the integrand in Eq. (\ref{eq:fdfsimple})
is underestimated at $z<1.25 a$ but it is overestimated at $1.25a<z<2.5a$ 
 (see Fig. \ref{fig:alpha_sphere}).
Using Eq. (\ref{eq:fdfapprox}) and following the same mathematical procedure 
as in Ostriker (1999),  we obtain $r_{\rm min}=0.71a$ for the homogeneous
sphere. In the case of a Plummer pontential with core radius $R_{s}$,
we infer $r_{\rm min} \simeq 1.36R_{s}$ at times $(\mach-1)c_{\infty}t\gg R_{s}$.
These estimates suggest that $r_{\rm min}$ is expected to depend weakly on
$\mach$, because what determines the region where the gravitational potential is no longer
described by a point-mass potential is $R_{s}$. Nevertheless, the goodness of the approximation 
for $F_{DF}$ given in Eq. (\ref{eq:fdfapprox}) 
can only be checked through explicit calculations. 

\begin{figure}
  \includegraphics[width=85mm,height=70mm]{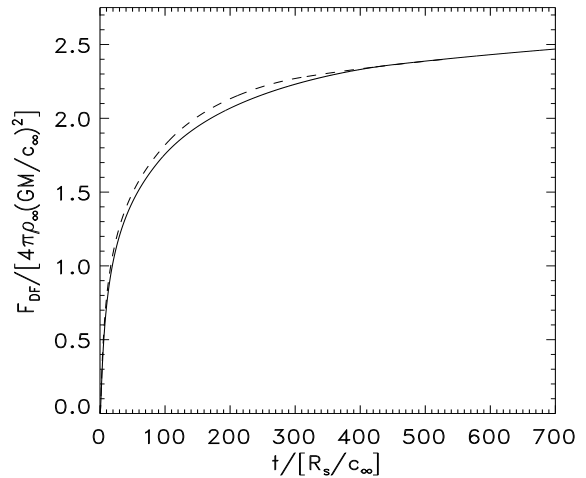}
  \caption{Dimensionless drag force, integrated on the whole computational
domain, versus time for a Plummer perturber with $\mach=1.5$
and $\anonlin=0.01$ for adiabatic (solid line) and isothermal gas (dashed line).
  }
  \label{fig:adi_iso}
\end{figure}

In the case where the perturber can be described by a Plummer sphere,
we have calculated $r_{\rm min}$ using three different approaches: using
a fully hydrodynamical code (S\'anchez-Salcedo \& Brandenburg 1999),
solving the time-dependent linear equations (S\'anchez-Salcedo 2012) and performing
the 3D integral defined in Equation (\ref{eq:alpha_extended}) 
numerically (S\'anchez-Salcedo 2009). 
We have always found $r_{\rm min}=2.25R_{s}$, which is slightly larger 
than the value inferred following Rein's approximation.  However, 
our value is significantly different from the value
inferred by KK09: $r_{\rm min}=0.35{\mathcal{M}}^{0.6}R_{s}$.
 For instance, at ${\mathcal{M}}=1.5$, the fitting formula given by
 KK09 predicts $r_{\rm min}=0.45R_{s}$. The difference in the values
 for $r_{\rm min}$ is remarkable. At a time $15R_{s}/V_{0}$, i.e. when the perturber has travelled
a distance of $15 R_{s}$, the drag force using $r_{\rm min}=0.45R_{s}$ is twice the drag force when
 a $r_{\rm min}$-value of $2.25R_{s}$ is used instead. Since the drag force in the linear regime is used as 
the reference value, it is important to have a robust determination of $r_{\rm min}$.

KK09 argued that the difference
 between the two prescriptions could be due in part to using different equations of
 state and in part to the lower spatial resolution used in S\'anchez-Salcedo \& Brandenburg
 (1999) simulations. None of these possibilities is fully satisfactory. In S\'{a}nchez-Salcedo
 \& Brandenburg (1999), we used a non-uniform grid with more resolution
 in the vicinity of the perturber, leading to $10$ zones per $R_{s}$. On the other hand, 
since the response of the gas is linear, Equations (\ref{eq:rephaeli})-(\ref{eq:ostriker2})
are valid when the gas is either 
 adiabatic or isothermal, as long as $\mach$ is taken as the
 ratio between the velocity of the perturber and the {\it relevant} sound speed
 in the medium. 

In order to explore the origin of this discrepancy, we decided to run adiabatic
simulations, using the same parameters, code and resolution as in KK09.
Simulations with higher resolution produced no appreciable change in the drag force.
We define the dimensionless drag force as $F_{DF}/F_{0}$ where
\begin{equation}
F_{0}=\frac{4\pi \rho_{\infty} (GM)^{2}}{c_{\infty}^{2}}.
\end{equation}
The dimensionless drag force as a function of Mach number at $t=10^{4}R_{B}/c_{\infty}=100R_{s}/c_{\infty}$,
for simulations with ${\mathcal{A}}=0.01$,
is shown in Figure \ref{fig:force_mach_lin}, together with the predicted values using Ostriker's formula 
(Eqs. \ref{eq:ostriker1} and \ref{eq:ostriker2}) for subsonic and supersonic perturbers, respectively.
In the supersonic cases, we used $r_{\rm min}=2.25R_{s}$, which corresponds to $\lambda=2.25$.
The agreement between simulations
and the analytical formula is very good. We also checked that the temporal evolution
of the drag force is fairly reproduced in these cases, using $r_{\rm min}=2.25R_{s}$. 
In order to make a more direct comparison with KK09 findings,
we explored a rather artificial case where $r_{\rm min}$ depends on the Mach number, as
suggested in KK09. We found that $r_{\rm min}=1.5 \mach^{0.6}R_{s}$ could
provide an acceptable fit to the evolution of the drag force for Mach numbers 
between $1$ and $4$, but only at times $>130 R_{s}/c_{s}$. Still, 
the derived value is a factor $4.3$ larger than the value quoted in KK09.

As an independent check, we ran the same simulations but assuming an
isothermal equation of state. As expected, the dimensionless drag
force is not sensitive to the adopted equation of state 
(see Figure \ref{fig:adi_iso}). 
 In the next section, we consider nonlinear cases.

\begin{figure*}
  \includegraphics[width=155mm,height=90mm]{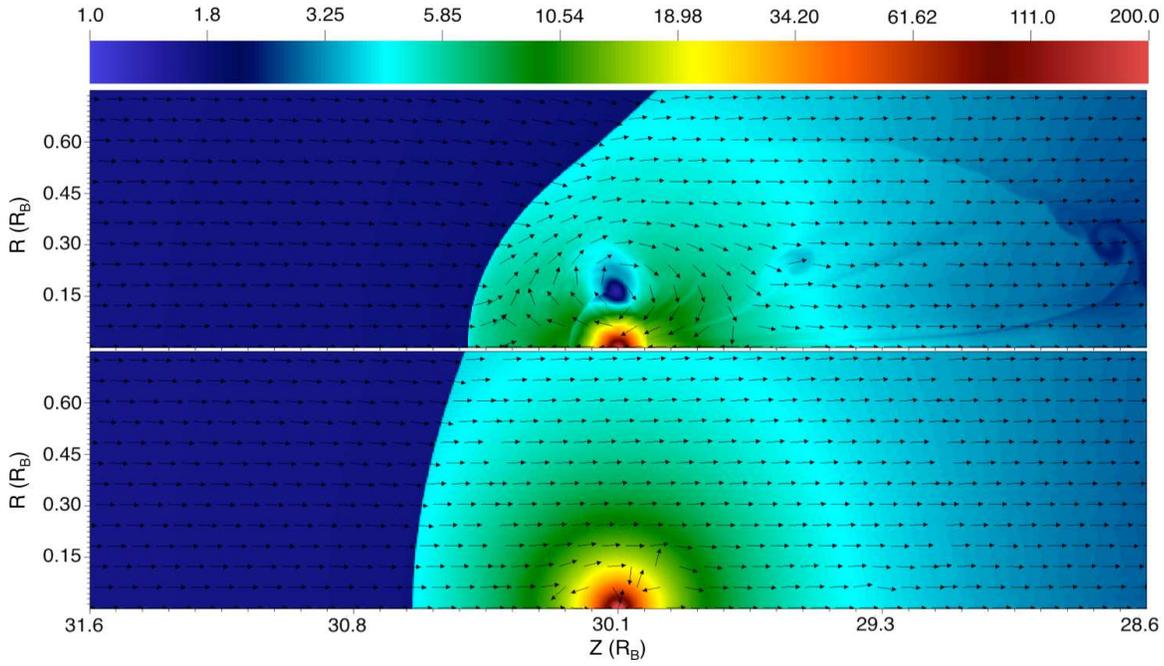}
\vskip 0.4cm
  \caption{Density color maps in logarithmic scale, and velocity field at $t=1.5R_{B}/c_{\infty}$  (upper panel)
and $t=15R_{B}/c_{\infty}$ (lower panel) for a gravitational body with $\anonlin=50$ and
$\mach=1.5$. The gas is adiabatic.
  }
  \label{fig:map_A50_M15}
\end{figure*}

\section{The nonlinear supersonic case: The values of $\tilde{\lambda}$}
\label{sec:nonlinear}
The complex and multidimensional dynamics of the flow in core potentials
has been studied in the adiabatic nonlinear case by KK09 and,
more recently, by Naiman et al.~(2011), including
also the isothermal cases. The latter authors,
however, were interested in the accumulation of mass around the center of the
potential and did not report on the drag force.
In order to have a reliable estimate of the drag force,
we carried out (adiabatic) numerical simulations varying $\anonlin$ between $8$ and
$50$, for various Mach numbers. 
In our simulations, we fixed $M$, $\rho_{\infty}$ and $c_{\infty}$ and only
${\mathcal{M}}$ and ${\mathcal{A}}$ vary from one simulation to another.
For $G=M=c_{\infty}=1$, the time unit is $R_{B}/c_{\infty}$ and so  
we define the dimensionless time as $\tilde{t}=(c_{\infty}/R_{B})t$.
In these units, given
${\mathcal{A}}$ and $\mach$, it holds that $R_{s}=1/{\mathcal{A}}$  and $V=\mach$.

We found that the dynamics of our simulated flow is consistent with those
performed in KK09 and Naiman et al. (2011).
For subsonic perturbers, we confirmed the result by KK09 that 
the nonlinear drag force reaches an asymptotic value that is similar 
to the drag 
in the linear case, as given in Eq. (\ref{eq:ostriker1}), regardless the value of $\anonlin$. Therefore,
we will not discuss the subsonic case any further; the reader interested
in the subsonic case is referred to KK09.

\begin{figure*}
  \includegraphics[width=160mm,height=90mm]{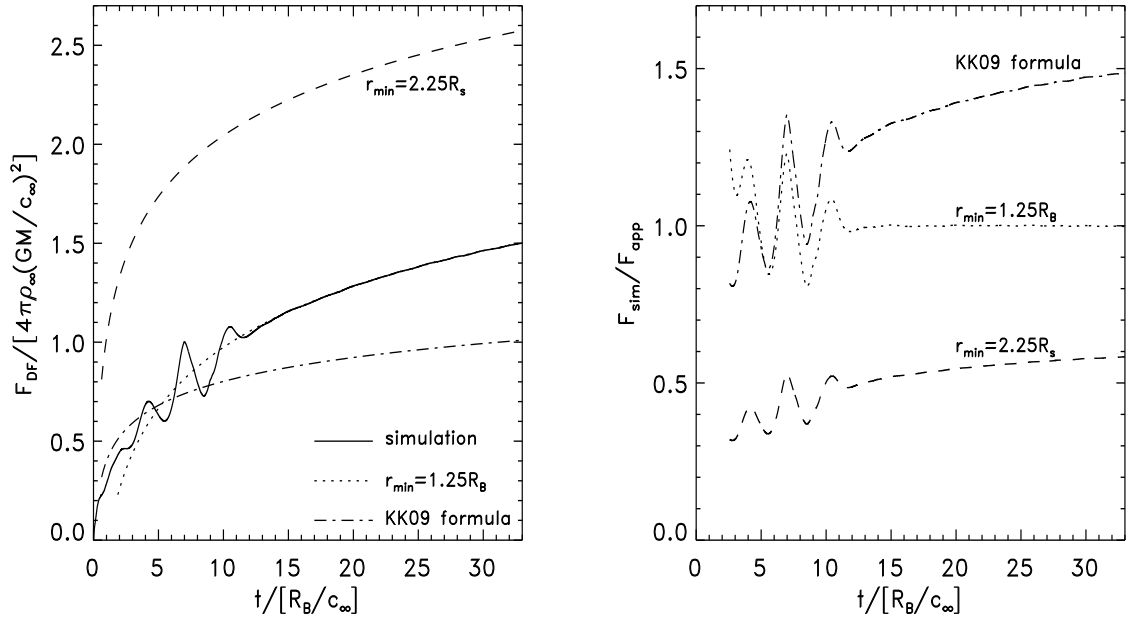}
\vskip 0.4cm
  \caption{Dimensionless drag force using three analytical formulae (Ostriker's formula 
[Eq. \ref{eq:ostriker2}] with $r_{\rm min}=2.25R_{s}$ [dashed line]; Eq. \ref{eq:forcedeepr} 
with $\tilde{\lambda}=1.25$ [implying $r_{\rm min}=1.25R_{B}$, dotted line];
and KK09 formula [Eq. \ref{eq:KK09formula}, dot-dashed line]),
as compared with the measured values from the simulation, for a body with $\mach=1.5$ and $\anonlin=20$.
The left-hand panel shows them as a function of time, whereas
the right-hand panel shows the ratio between the values obtained in the simulation $F_{\rm sim}$
and the values when the abovementioned analytical equations are used.
  }
  \label{fig:force_time_M15}
\end{figure*}

\begin{figure}
  \includegraphics[width=90mm,height=80mm]{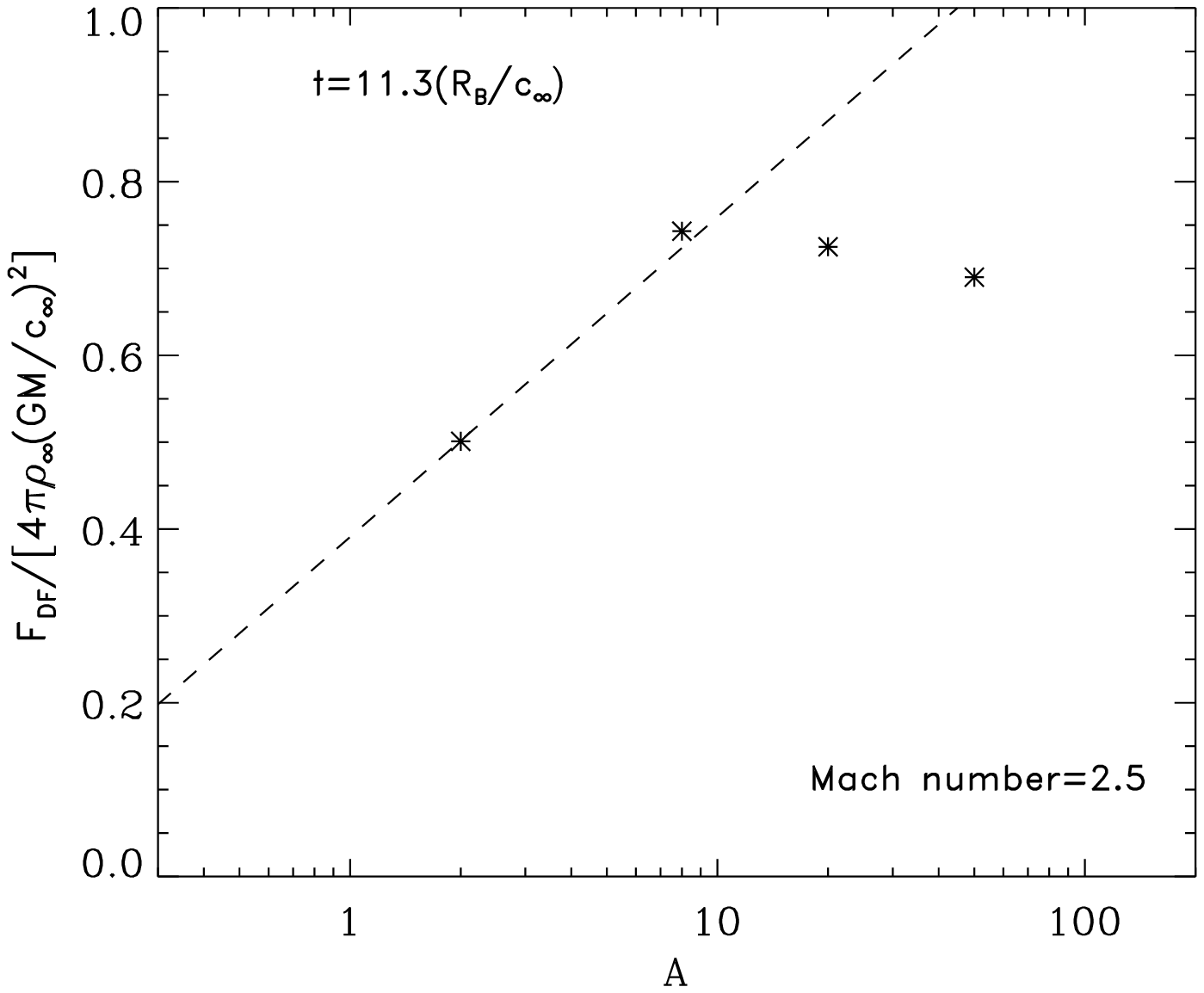}
\vskip 0.4cm
  \caption{Dimensionless drag force at $\tilde{t}=11.3$, as a function of $\anonlin$ for
a perturber with $\mach=2.5$ (asterisks). The dashed line represents the drag force derived in the linear
approximation (Equation \ref{eq:flin_A} with $\lambda=2.25$).
  }
  \label{fig:force_A_Mach25}
\end{figure}

For supersonic perturbers with ${\mathcal{A}}\gtrsim 1$, the linear analysis cannot capture the flow dynamics
in regions with high spatial gradients; that is, 
on the surface of the Mach cone and at distances $r\lesssim R_{B}$ from the perturber.
In Figure \ref{fig:map_A50_M15}, we plot the density color maps and the velocity field of the flow around
a perturber with $\mach=1.5$ and $\anonlin=50$.
As already described by KK09, a bow shock thermalizes the
upstream flow, leading to the formation of an envelope of highly subsonic gas in
quasi-hydrostatic equilibrium. In fact,
since the flow at infinity is laminar, shocks may
occur for those streamlines that are bent significantly.
Streamlines with impact parameters $b$ obeying
the condition
\begin{equation}
\frac{2GM}{\sqrt{b^{2}+R_{s}^{2}}}<c_{\infty}^{2}+V_{0}^{2},
\end{equation}
will bend significantly and terminate in a shock (e.g., Krumholz et al. 2005). 
In terms of $R_{A}$, this condition for strong bending can be written as
\begin{equation}
b<\left(R_{A}^{2}-R_{s}^{2}\right)^{1/2}.
\end{equation}
In a case where ${\mathcal{A}}\gg 1$, then
$R_{s}\ll R_{A}$, condition that reduces to $b<R_{A}$.
The incoming supersonic flow moving close to the axis of symmetry collides
with streamlines that have larger impact parameters but have been bent by
the gravitational potential, leading to the formation of a detached bow shock.
We find that for $\mach=1.5$ and $\anonlin=50$, the stand-off distance of the
bow shock to the perturber along the symmetry axis ($R=0$),
denoted by $\delta$, is $\sim 0.56R_{B}=0.9R_{A}$, which
is in good agreement with the value reported in KK09 (see their Figure 12).
Since the contribution to the drag force of material within a sphere of radius $\sim \delta$ is very small,
because of the front-back symmetry, 
we expect $r_{\rm min}\gtrsim R_{A}$, which implies
$\tilde{\lambda}\gtrsim 2/(1+\mach^{2})$ in the deep nonlinear regime (remind that $r_{\rm min}=\tilde{\lambda}R_{B}$).

In a situation where $R_{A}\gg 2.25R_{s}$ (which also implies that $R_{B}\gg 2.25R_{s}$),
one expects that the relevant minimum radius $r_{\rm min}$ for the
gravitational interaction will be solely determined by
$R_{B}$ and $\mach$, because $R_{s}$ becomes
irrelevant as far as the drag force concerns.
Under these circumstances, the drag force, as a function of $\tilde{t}$,  is expected 
to be independent of $\anonlin$:
\begin{equation}
\frac{F_{DF}}{F_{0}}=\frac{1}{\mach^{2}}
\left[\frac{1}{2}\ln \left(1-\mach^{-2}\right)+
\ln\left(\frac{\mach\tilde{t}}{\tilde{\lambda}}\right)\right].
\label{eq:forcedeepr}
\end{equation}
The parameter $\tilde{\lambda}$ that connects $r_{\rm min}$ with $R_{B}$ 
may be a complex function of the Mach number $\mach$, but it should also
depend on the equation of state  (i.e. on $\gamma$ for polytropic gas).
The condition $R_{A}\gg 2.25 R_{s}$ can be written in terms of a critical
value ${\mathcal{A}}_{\rm cr}$,
as $\anonlin \gg \anonlin_{\rm cr}$, with $\anonlin_{\rm cr}=\mach^{2}+1$.
For a typical value of $\mach$, say $\mach=2$, $\anonlin_{\rm cr}= 5$. 

\begin{figure}
  \includegraphics[width=90mm,height=130mm]{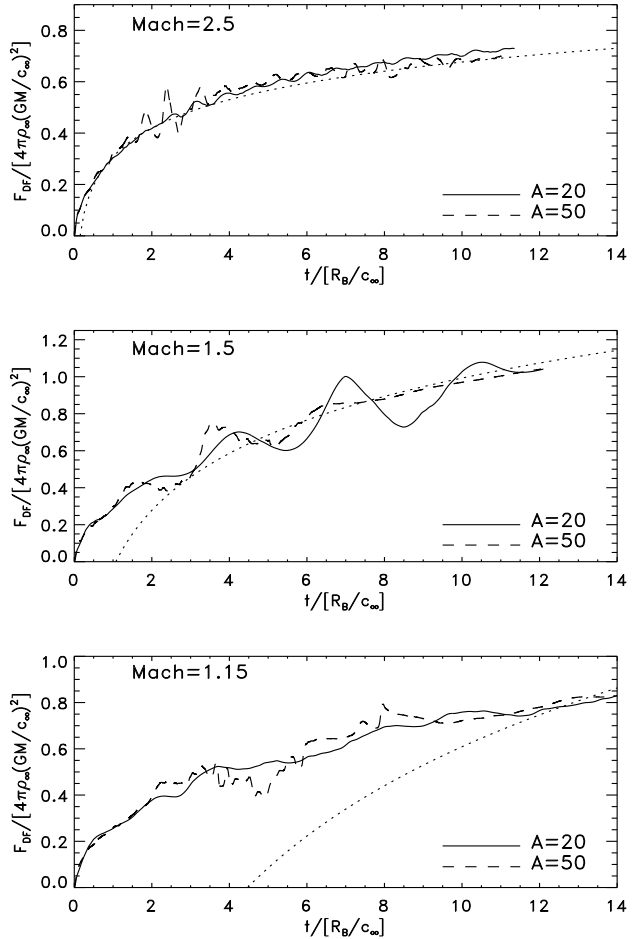}
\vskip 0.4cm
  \caption{Drag force as a function of time for $\anonlin=20$ (solid lines) and 
$\anonlin=50$ (dashed lines). The correponding Mach number is quoted at the
corner of each panel. The dotted curves give the drag force using Equation (\ref{eq:forcedeepr})
with $\tilde{\lambda}=0.33$ (top panel),
$\tilde{\lambda}=1.20$ (middle panel) and $\tilde{\lambda}=2.54$ (lower panel).
  }
  \label{fig:force_time_all}
\end{figure}

Figure \ref{fig:force_time_M15} shows the drag force as a function of $\tilde{t}$ for a body moving
at $\mach=1.5$ and $\anonlin=20$ (thus, $\anonlin>\anonlin_{\rm cr}=3.25$). 
We see that, at $\tilde{t}>3$, Eq. (\ref{eq:forcedeepr}) 
with $\tilde{\lambda}=1.25$ (implying that $r_{\rm min}=1.25 R_{B}$) reproduces
the mean amplitude (not the fluctuations) and temporal evolution of the drag force fairly well. 
For comparison, we also plot $F_{\rm lin}$, that is,
the drag force using Ostriker's formula (Eqs. \ref{eq:rephaeli} and \ref{eq:ostriker2}) with
$r_{\rm min}=2.25R_{s}$.
If the time is measured in units of $R_{B}/c_{\infty}$, $F_{\rm lin}$ is given by
\begin{equation}
\frac{F_{\rm lin}}{F_{0}}=\frac{1}{\mach^{2}}
\left[\frac{1}{2}\ln(1-\mach^{-2})+\ln\left(\frac{{\mathcal{A}} \mach\tilde{t}}{\lambda}\right)\right],
\label{eq:flin_A}
\end{equation}
with $\lambda\simeq 2.25$ (see Section \ref{sec:linearcase}).
As expected, a value of $r_{\rm min}=2.25R_{s}$ overestimates the drag force at
any time.

KK09 suggested an empirical relation between $F_{DF}$ and $F_{\rm lin}$
through the parameter $\eta\equiv \anonlin/(\mach^{2}-1)$. They found that 
\begin{equation}
F_{DF}=(\eta/2)^{-0.45}F_{\rm lin},
\label{eq:KK09formula}
\end{equation} 
for $2\leq \eta\leq 100$. Figure \ref{fig:force_time_M15} shows that
the ratio $F_{DF}/F_{\rm lin}$ is not  constant with time; its value
is always larger than $(\eta/2)^{-0.45}=0.39$, implying that KK09 prescription
underestimates the drag force. The ratio $F_{DF}/F_{\rm lin}$ is not constant with time but approaches
asymptotically to $1$ when $\tilde{t}\rightarrow \infty$, i.e. when the
contribution of the inner part of the wake to the gravitational drag becomes small
as compared to the contribution of the far-field wake, which increases logarithmically
with time.
In fact, at times
$t\gg R_{B}/V_{0}$, the nonlinear part of the wake has reached a quasistationary
regime; $F_{DF}$ increases with time due to the new material that is 
incorporated
into the wake in the far field, which can be described in linear theory.
This implies that $dF_{DF}/dt=dF_{\rm lin}/dt$.
Therefore, at large enough times, what is constant with time is the difference
$F_{\rm lin}-F_{DF}=F_{0}\ln(\tilde{\lambda}R_{B}/[\lambda R_{s}])=
\ln(\tilde{\lambda}\anonlin/\lambda)$.  
The formula $F_{DF}=(\eta/2)^{-0.45}F_{\rm lin}$ may induce a significant fractional error 
at large Coulomb logarithms.

Figure \ref{fig:force_A_Mach25} shows the drag force at $\tilde{t}=11.3$ for a perturber with $\mach=2.5$
and differing ${\mathcal{A}}$. Note that $\anonlin_{\rm cr}=7$
in this case. We see that the drag force when ${\mathcal{A}}<8$
increases linearly with $\ln{\mathcal{A}}$, as predicted in linear theory (see Eq. \ref{eq:flin_A}). 
For ${\mathcal{A}}>8$,
the drag force is essentially constant with ${\mathcal{A}}$. The transition between
the linear regime and the plateau, where the force becomes almost constant, is abrupt. 

\begin{figure*}
  \includegraphics[width=155mm,height=90mm]{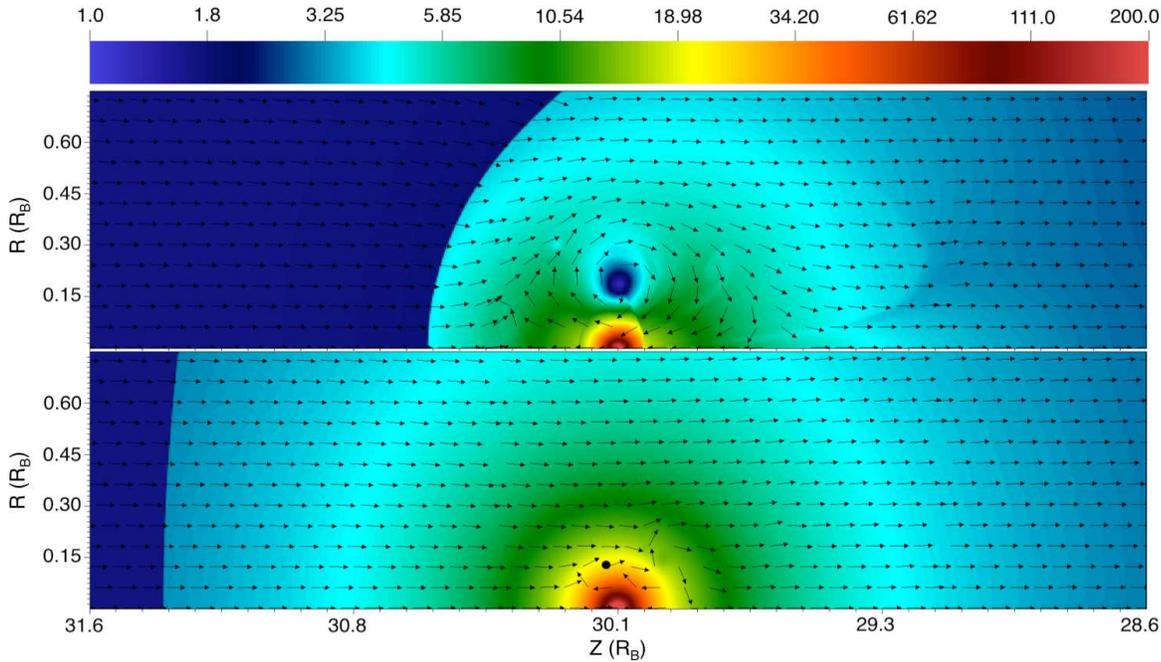}
\vskip 0.4cm
  \caption{Density color maps in logarithmic scale and velocity field at $t=1.5R_{B}/c_{\infty}$  (upper panel)
and $t=15R_{B}/c_{\infty}$ (lower panel) for a gravitational body with $\anonlin=50$ and
$\mach=1.15$. The gas is adiabatic.
  }
  \label{fig:map_A50_M115}
\end{figure*}

Figure \ref{fig:force_time_all} shows the drag force, as a function of time, for $\mach=2.5, 1.5$, and $1.15$,
with $\anonlin=20$, and $50$. The drag force displays some oscillations with time, but the
amplitudes of the drag force for $\anonlin=20$ and $\anonlin=50$ are similar. At Mach numbers
$1.5$ and $2.5$, Equation (\ref{eq:forcedeepr}) predicts the amplitude of the drag force correctly for
$\tilde{\lambda}=0.33$ and $\tilde{\lambda}=1.2$, respectively. This clearly shows that
$\tilde{\lambda}$ depends strongly on $\mach$. At Mach number $1.15$, a $\tilde{\lambda}$-value
of $2.54$ is required to fit the amplitude of the drag force at $\tilde{t}=12$--$14$, but it fails
to reproduce the strength of the drag force at shorter times. In fact, as the Mach number approaches
to $1$, an increasingly longer time is required until the drag force increases logarithmically in time.
Thus, Equation (\ref{eq:forcedeepr}) is not a good approximation at early times in those cases. 
The reason is that perturbers moving at speed near Mach $1$,  can sustain a large envelope of gas, and
the time required for the perturber to establish such a quasi-stationary density envelope increases
when $\mach\rightarrow 1$. This is illustrated in Figure \ref{fig:map_A50_M115}. We see the temporal evolution
of the bow shock and the formation of the envelope for a perturber moving at $\mach=1.15$,
and with $\anonlin=50$. For this transonic Mach number, the distance of the
bow shock at $\tilde{t}=15$ is a factor of $2$ larger than for $\mach=1.5$.
This explains why the minimum radius is larger 
at $\mach=1.15$ than it is at $\mach=1.5$. 

\begin{figure}
  \includegraphics[width=90mm,height=80mm]{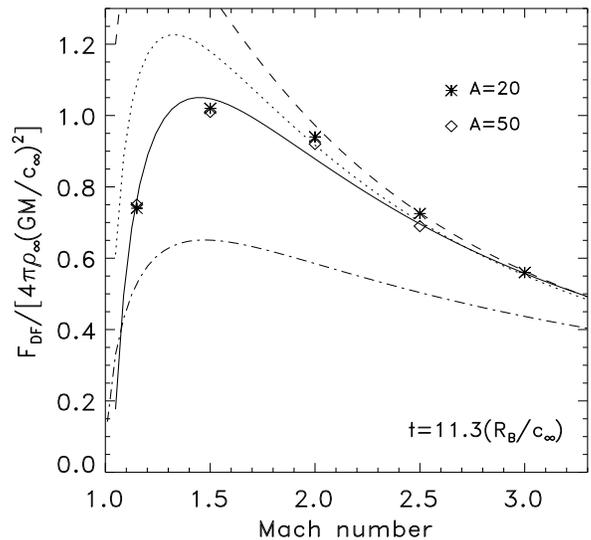}
\vskip 0.4cm
  \caption{Measured dimensionless drag force at $\tilde{t}=11.3$, as a function of $\mach$,
for $\anonlin=20$ (asterisks) and $\anonlin=50$ (diamonds). The lines shown are the
predicted values using Equation (\ref{eq:forcedeepr}) with $\tilde{\lambda}=3.3\mach^{-2.5}$
(solid line), with $\tilde{\lambda}=2\mach^{-2}$ (dotted line) and with $\tilde{\lambda}=2/(\mach^{2}+1)$
(dashed line). The drag force using Equation
(\ref{eq:KK09formula}) with $\anonlin=50$ is also shown (dot-dashed line).
 }
  \label{fig:force_Mach_11}
\end{figure}

Figure \ref{fig:force_Mach_11} shows the drag force at $\tilde{t}=11.3$, as a function of Mach number. 
As reported in KK09, the Mach number corresponding to the maximum drag
force shifts from unity to $\sim 1.5$ as the wake becomes highly nonlinear. A minimum
radius with $\tilde{\lambda}=3.3\mach^{-2.5}$ provides a good fit to the drag force for Mach
numbers between $1.5$ and $3$. As already said, at Mach numbers close to $1$, the value of $\tilde{\lambda}$
cannot be established because, at those times, Eq. (\ref{eq:forcedeepr}) does not provide
a good description of the temporal behaviour of $F_{DF}$. It is important to note that
at Mach numbers $\geq 2$, $\tilde{\lambda}=2/(\mach^{2}+1)$ provides a good fit, but especially
a power law $\tilde{\lambda}=2.0\mach^{-2}$, implying
that $r_{\rm min}=2GM/V_{0}^{2}$, gives a better fit.

It is interesting to consider the point-mass case as the limit when $R_{s}=0$.
The structure of the wake past a body with a nonzero $R_{s}$ is
different to a case where $R_{s}=0$. In the first case, there is conservation of mass in the whole 
computational domain, whereas in the second case one must implement an inner boundary condition
where ``accreted'' gas particles are sinked. Cant\'o et al. (2011) studied the 
point-mass case in the hypersonic case and found $r_{\rm min}\simeq 0.8GM/V_{0}^{2}$. Therefore, although the flow
past a point-mass accretor is different to the flow past a non-accretor, the minimum impact
parameter in the drag force might be relatively similar. 

\section{Discussion and conclusions}
\label{sec:conclusions}
The interaction between a gravitational perturber and the ambient medium is a classical problem
in astrophysics (e.g., Bondi \& Hoyle 1944). Star clusters moving in cold interstellar gas, 
globular clusters in gaseous halos,
galaxies moving through the intergalactic or intracluster medium, can be described as core potentials.
In these systems, it is usually assumed that their individual members are accreting at so low enough
rates that the mass is conserved (i.e. gas removal due to accretion can be neglected, see e.g., Naiman et al. 2011).
In this paper, we have studied the gravitational drag force (dynamical friction) that their 
own induced wake produces on this kind of objects. To do so, we have modelled them as Plummer
perturbers and the gas is assumed adiabatic. We have studied both the linear and 
nonlinear regimes.

In the supersonic linear regime, the maximum of the mass density in the wake created by
a core (non-diverging) potential is not situated at the minimum
of the gravitational potential but it is placed behind it. In the linear regime, the minimum radius $r_{\rm min}$
cannot be much larger than a few softening radii because beyond those distances from
the perturber, the wake density is almost undistinguishable from the wake created by a point mass.
In particular, for the Plummer sphere with core radius $R_{s}$, we confirm
our previous result that the minimum
radius for the drag force is $\sim 2.25 R_{s}$.
Thus, $r_{\rm min}$ is fairly
independent of the Mach number. We must emphasize that in the linear regime, 
the dependence on the particular
radiative cooling and heating of the gas is through the sound speed $c_{\infty}$.

If we keep $R_{B}$ fixed and reduce $R_{s}$, we have models with higher $\anonlin$ and, hence,
the gas dynamics around the body becomes nonlinear.
When $R_{B}\gg R_{s}$ (i.e. $\anonlin\gg 1$), the radius $R_{s}$ becomes irrelevant
as far as the drag force is concerned. Thus, the minimum radius $r_{\rm min}$ must be related 
with the Bondi scale and/or the ``accretion'' radius.
Using axisymmetrical simulations, we found that the drag force can be described
well by using Ostriker (1999) formula with $r_{\rm min}=3.3\mach^{-2.5}R_{B}$ for perturbers
with $\mach\geq 1.5$ and $\anonlin>\anonlin_{\rm cr}\equiv\mach^{2}+1$. However,
 at $\mach>2$, the drag force can be equally explained using $r_{\rm min}=2\mach^{-2}R_{B}$
or $r_{\rm min}=R_{A}$. The latter possibilities imply that $r_{\rm min}=2GM/V_{0}^{2}$
at the hypersonic limit, which
resembles the scaling found for a point-mass (accreting) perturber by Cant\'o et al. (2011).

At large enough times, the wake can be decomposed into the nonlinear part, which
has reached a quasi-stationary state and a growing far-field wake. 
Our working formula for the drag force naturally includes the fact that at times $\gg R_{B}/c_{\infty}$,
the material that is being incorporated into the wake has large impact parameters and thus
can be described using the linear theory, implying that 
$dF_{DF}/dt=dF_{\rm lin}/dt$. On the contrary, the formula proposed by KK09,
$F_{DF}=(\eta/2)^{-0.45}F_{\rm lin}$,
does not match this condition, meaning that this relationship between $F_{DF}$ and $F_{\rm lin}$
should break down at a certain $t$. We tested KK09 formula and 
found that it underestimates the drag force, 
especially at large times (long wakes). 

Applied to a perturber in orbit,
KK09 argue that since $F_{DF}\propto M^{1.55}$, then
the orbital decay timescale of a cored perturber declines as 
$\tau_{\rm dec}\propto M^{-0.55}$ (e.g., Binney \& Tremaine 1987). Our formula has the classical
dependence $F_{DF}\propto M^{2}\ln \Lambda$, where the dependence
of $\ln\Lambda$ on $M$ is very weak. Thus, perturbers with an initial orbital radii
much larger than $R_{B}$ decay in a characteristic timescale $\propto M^{-1}$.

\acknowledgments
We are grateful to DGTIC-UNAM for allowing us to use its KanBalam and
Miztli Clusters, where all the simulations were performed. 
Some preliminary simulations were carried out using Atocatl at IA-UNAM.
The software used in this work was in part developed by the 
DOE NNSA-ASC OASCR Flash Center at the University of Chicago.
We acknowledge financial support from CONACyT project 165584
and PAPIIT project IN106212.

\end{document}